\documentclass[10pt,twocolumn,letterpaper]{article}

\usepackage[pagenumbers]{cvpr} 

\definecolor{cvprblue}{rgb}{0.21,0.49,0.74}
\usepackage[pagebackref,breaklinks,colorlinks,allcolors=cvprblue]{hyperref}

\def\paperID{1} 
\def\confName{CVPR}
\def\confYear{2025}

\title{Aligning Proteins and Language: A Foundation Model for Protein Retrieval}

\author{
Qifeng Wu$^{1}$\thanks{Accepted to the CVPR 2025 Workshop on MMFM-BIOMED.}, Zhengzhe Liu$^{1}$, 
Han Zhu$^{2}$, Yizhou Zhao$^{1}$,\\
Daisuke Kihara$^{2}$, Min Xu$^{1}$\thanks{Corresponding author. This work was supported in part by the U.S. NSF grant IIS-2211597.}\\[3pt]
$^1$Carnegie Mellon University, $^2$Purdue University
}

\usepackage{float}
\usepackage{caption}
\usepackage{subcaption}
\usepackage{adjustbox}
\usepackage{float}
\usepackage{caption}
\usepackage{subcaption}
\usepackage{booktabs}
\usepackage[dvipsnames]{xcolor} 
\newcommand\minisection[1]{\vspace{1mm}\noindent \textbf{#1}}

\begin{document}
\maketitle
\begin{abstract}

This paper aims to retrieve proteins with similar structures and semantics from large-scale protein dataset, facilitating the functional interpretation of protein structures derived by structural determination methods like cryo-Electron Microscopy (cryo-EM). Motivated by the recent progress of vision-language models (VLMs),  
we propose a CLIP-style framework for aligning 3D protein structures with functional annotations using contrastive learning. For model training, we propose a large-scale dataset of approximately 200,000 protein-caption pairs with rich functional descriptors. We evaluate our model in both in-domain and more challenging cross-database retrieval on Protein Data Bank (PDB) and Electron Microscopy Data Bank (EMDB) dataset, respectively. In both cases, our approach demonstrates promising zero-shot retrieval performance, highlighting the potential of multimodal foundation models for structure-function understanding in protein biology.
\end{abstract}    
\vspace{-15pt}
\section{Introduction}
\label{sec:intro}

Understanding protein function from structural data is a central goal in computational biology, with applications spanning drug discovery, enzyme engineering, and systems biology. Recent advancements, particularly in Cryo-EM, have dramatically facilitated protein 3D structural modeling; yet, efficiently retrieving functionally similar proteins based on structural data remains challenging, limiting our ability to leverage the exponentially growing structural databases for functional annotation and knowledge transfer.

In this paper, we introduce a foundation protein retrieval model to bridge the gap between 3D protein structures, such as those captured by Cryo-EM, and their associated functional descriptions in natural language. Specifically, given a query protein structure, the model aims to accurately retrieve functionally related proteins and relevant textual information in a zero-shot manner.


The retrieval task has been widely studied in a wide range of areas ranging from natural images to biomedical domains.
Recent advances in CLIP-style models~\cite{radford2021learning,li2021supervisio} have demonstrated remarkable success in zero-shot image retrieval 
across large-scale datasets. 
Besides, several works have explored multi-modal pretraining for retrieval in biomedical domains, including radiology~\cite{zhou2022generalized, lee2024read, yildirim2024multimodal}, pathology~\cite{qiao2022multi, dai2025pathologyvlm, zeng2024transferring}, and electronic health records~\cite{boag2018cliner,rahman2024next,alsaad2024multimodal}. However, vision-language-based protein and function alignment remain underexplored. To fill this gap, this work focuses on protein structure retrieval by training a multi-modal foundation model that connects proteins and text annotations through contrastive learning.

Multi-modal retrieval of protein structures presents unique challenges. First, how to encode and represent the observed 3D protein structures remains unsolved due to its high-resolution and spatially complex details. 
Functional annotations, meanwhile, are hierarchical and semantically dense, challenging to describe thoroughly. 
Second, prior work in structure-function prediction often relies on supervised classifiers~\cite{gligorijevic2021structure} trained on a limited manually annotated dataset, 
limiting its scalability and generalization. 
Third, robust cross-dataset generalization is crucial. This capability allows a model trained on a large-scale dataset to effectively retrieve relevant structures or annotations even when dealing with data from different sources or resolutions.

To address these issues, we develop a multi-modal foundation model, allowing relevant protein structure or annotation retrieval through a large protein database. First, we construct a large-scale dataset consisting of around 200,000 protein-text pairs for model pretraining. In our dataset, protein samples cover a wide range of samples of PDB dataset and text annotations contains rich semantic functional information.  
Second, We develop a CLIP-style model~\cite{radford2021learning} to bridge the protein structures and text annotations. The model is pretrained with contrastive learning objectives on our curated dataset to bring structurally and functionally aligned protein–caption pairs closer in the shared embedding space, and in the meantime, pushing apart mismatched or semantically unrelated pairs. 

We evaluate the zero-shot retrieval performance of our model, which achieves around 60\% Top-5 accuracy, demonstrating the effectiveness of our model in the challenging zero-shot setting. Furthermore, to examine the limit of our model and explore real-world retrieval scenario of an observed protein structure, we evaluate the \textbf{cross-database zero-shot retrieval} performance on EMDB entries which are not exposed to the model during pretraining. Again, our model achieves promising performance. 

\noindent We summarize our contribution in this work as follows:
\begin{itemize}
    \item \textbf{Task:} We explore a novel multi-modal protein retrieval task, facilitating the understanding of observed protein structures on which we have limited knowledge. 
    \item \textbf{Dataset:} We construct a large-scale dataset consisting of around 200,000 proteins and associated rich semantic-meaning annotations. 
    \item \textbf{Model:} We develop and pretrain a CLIP-style model~\cite{radford2021learning} to bridge the protein structures and text annotations for protein retrieval. 
    \item \textbf{Experimental results:} Our model demonstrates encouraging capabilities on zero-shot and zero-shot cross-domain protein retrieval tasks. 
\end{itemize}






\vspace{-4pt}
\section{Datasets and Preprocessing}
\label{sec:data}

\begin{figure*}
  \centering
    \includegraphics[width=0.9\linewidth]{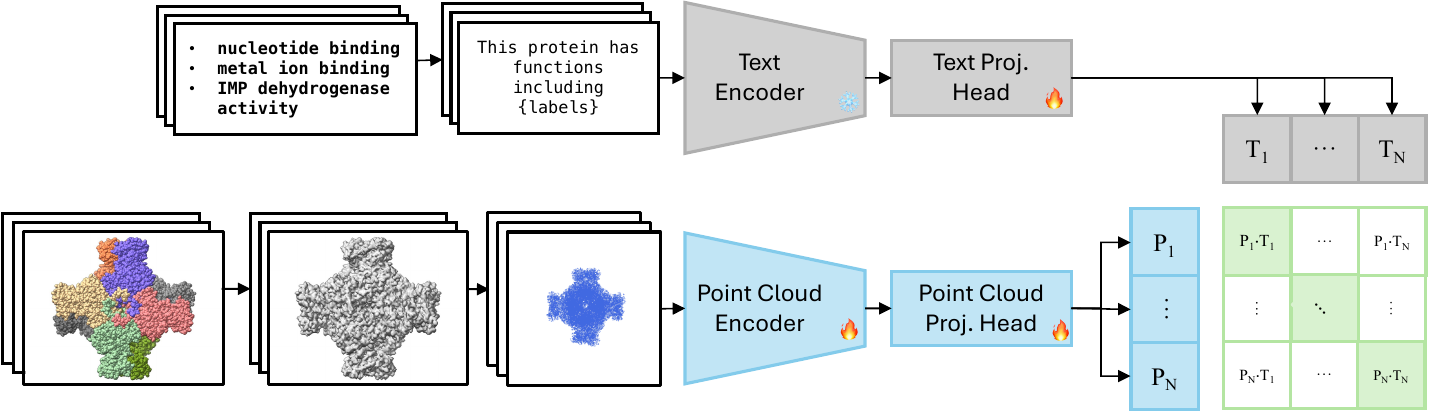}
  \caption{\textbf{Contrastive pretraining with protein–text pairs.} Protein structures are first converted to volumetric density maps synthesized from atomic coordinates (PDB), from which surface meshes are extracted via marching cubes and uniformly sampled into point clouds. Text descriptions are generated by concatenating Gene Ontology (GO) terms filtered for specificity and prepended with a natural language prompt. The \textcolor{orange}{fire} icon indicates trainable components, while the \textcolor{cyan}{Snow} icon denotes frozen modules.} 
  \label{fig:main}
\end{figure*}

\begin{figure}[t]
  \vspace{-8pt}
  \centering
  \includegraphics[width=\linewidth]{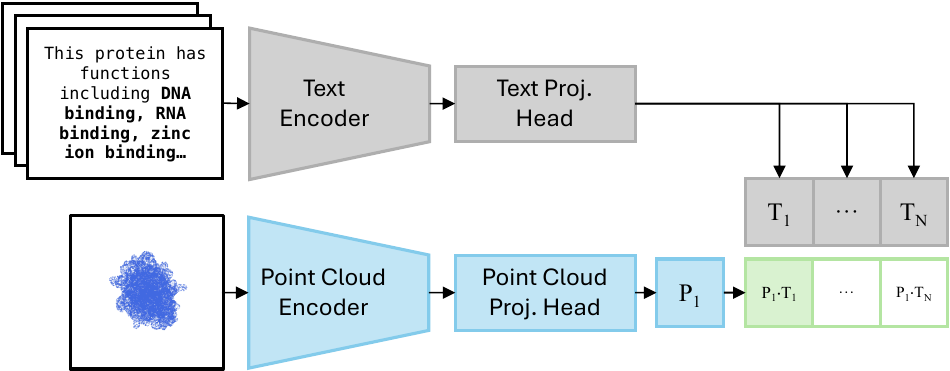}

  \caption{\textbf{Zero-shot retrieval.} Given a query protein point cloud, the model ranks a pool of candidate captions by similarity in the shared embedding space.}
  \label{fig:eval}
  \vspace{-12pt}
\end{figure}

\subsection{Data Sources}
Our dataset consists of protein samples sourced from the Protein Data Bank (PDB), paired with corresponding functional annotations from the Gene Ontology (GO). To evaluate cross-database retrieval, we additionally include entries from the Electron Microscopy Data Bank (EMDB) that are explicitly linked to PDB entries, enabling structural comparisons across different resolution regimes.
\subsection{Preprocessing Pipeline}
\minisection{PDB.}
We adopt point clouds as the structural representation of proteins due to their ability to provide a compact yet expressive encoding of 3D molecular geometry. In contrast to voxel grids, which tend to be sparse and computationally demanding, point clouds offer a memory-efficient alternative that effectively captures both the global conformation and fine surface features of proteins. This representation aligns naturally with modern geometric deep learning architectures and facilitates rotationally invariant learning—an essential property for modeling proteins with arbitrary orientations in 3D space. To derive geometric representations from PDB entries, we first convert atomic coordinates into volumetric density maps using the EMAN2 tool \texttt{pdb2mrc}~\cite{tang2007eman2}, simulating electron scattering to produce 3D density volumes. We then apply the marching cubes algorithm~\cite{lorensen1998marching} to extract surface meshes from the volumes. The iso-value, which determines the threshold density level at which a surface is extracted, is set empirically as:
\begin{equation}
\text{iso-value} = \mu + 0.5 \sigma,
\end{equation}
where $\mu$ and $\sigma$ denote the mean and standard deviation of the voxel intensities, respectively. This ensures consistent surface extraction across varying density distributions, ensuring consistent surface extraction across varying density distributions. After surface extraction, we uniformly sample 16,384 points to construct a standardized point cloud representation suitable for downstream encoding.

\minisection{Annotation.}
The functional annotation of each protein is represented as a caption composed of Gene Ontology (GO) terms. To construct these captions, we concatenate GO terms associated with the protein into a single sentence. Given the maximum input length constraint of the text encoder and the hierarchical nature of GO annotations, we apply a filtering strategy to improve specificity and reduce redundancy. In particular, we remove high-level or generic terms that have child terms and 
keep only the most specific leaf-level terms while ignoring others. (e.g., given the list of terms: \texttt{binding}, \texttt{DNA binding}, \texttt{RNA binding}, we remove the parent term \texttt{binding} and keep the rest.)
This ensures that our text descriptions are both compact and informative, preserving fine-grained informative functional descriptions while avoiding redundancy. 

\minisection{EMDB.}
For EMDB entries, which contain experimentally derived volumetric density maps, we apply similar procedures, marching cubes~\cite{lorensen1998marching} and uniform sampling, to extract point clouds, ensuring representation consistency with the PDB-derived samples. In this case, the iso-value used for surface extraction is set to the author-recommended contour level specified in the EMDB metadata. This empirically chosen threshold reflects the confidence level of the electron density and helps produce biologically meaningful surface meshes. Using overly low iso-values may result in noisy or bloated surfaces that include background artifacts, while overly high values may discard relevant structural regions. Adopting the recommended contour level provides a standardized and interpretable basis for mesh generation across entries.

With the preprocessing strategies introduced above, in our dataset, each protein sample is paired with a list of GO terms annotating its functions.





\vspace{-4pt}
\section{Method}
Our goal is to learn a shared embedding space where the representation of proteins and functional descriptions are aligned via contrastive learning.

\subsection{Model Architecture}

Inspired by CLIP~\cite{radford2021learning}, we adopt a dual-encoder architecture to embed protein structures and GO-based captions into a shared representation space. As shown in Fig. \ref{fig:main}, the model consists of a point cloud encoder $\phi_{\text{pc}}$ and a text encoder $\phi_{\text{text}}$, each followed by a projection head consists of linear, normalization, activation, and dropout layers.

To encode protein represented by point clouds, we adopt I2P-MAE~\cite{zhang2023learning}, which employs pretrained 2D transformer based vision model to supervise 3D representation learning, as the point cloud encoder and used the pretrained checkpoint released in Point-bind~\cite{guo2023point}. 

To encode text, we use the \texttt{all-mpnet-base-v2} model from the SentenceTransformers library~\cite{reimers2019sentence,song2020mpnet}, a pre-trained transformer-based encoder optimized for sentence-level semantic similarity and feature extraction. Captions derived form Gene Ontology are input as raw text and tokenized using the standard MPNet tokenizer.

Then, we pass features obtained from each encoder through separate linear projection heads to produce embeddings in a shared embedding space:
\vspace{-5pt}
\begin{equation}
P = W_{\text{pc}} \phi_{\text{pc}}(x), \quad T = W_{\text{text}} \phi_{\text{text}}(t),
\end{equation}
\vspace{-12pt}

where $W_{\text{pc}}$ and $W_{\text{text}}$ are learned projection matrices. The retrieval is cosine similarity-based, as shown in~\cref{fig:eval}. 

\subsection{Training Objective}

As shown in~\cref{fig:main}, the model is trained using a symmetric InfoNCE loss over a batch of $N$ protein–caption pairs:
\begin{equation}
\mathcal{L} = \frac{1}{2N} \sum_{i=1}^{N} \left[ \mathrm{CE}(s_{i,:}, y_i) + \mathrm{CE}(s_{:,i}, y_i) \right],
\end{equation}
where $\mathrm{CE}(\cdot, y_i)$ denotes the cross-entropy loss with the $i$-th item as the correct label. $s_{i,:}$ and $s_{:,i}$ denote the $i$-th row and $i$-th column of the similarity matrix $s$ respectively. Particularly, the similarity matrix consists of pair-wise cosine similarities scaled by a learnable temperature parameter $\tau$:
\vspace{-6pt}
\begin{equation}
s_{i,j} = \frac{1}{\tau}\frac{P_i \cdot T_j}{\lVert P_i \rVert \lVert T_j \rVert}.
\end{equation}

This bidirectional objective aligns both modalities simultaneously, encouraging aligned pairs to be close in the embedding space while driving mismatched pairs apart.


\subsection{Training Details}

The model is trained for 40 epochs using the AdamW optimizer with a $1 \times 10^{-8}$ weight decay. The initial learning rate is set to $1 \times 10^{-3}$ and decayed using cosine annealing. We use a global batch size of 64, distributed across 4 NVIDIA A40 GPUs with a batch size of 16 per GPU.
During training, we freeze the text encoder (\texttt{all-mpnet-base-v2}) and optimize only the point cloud encoder (I2P-MAE) and the projection layers. This strategy leverages the strong semantic representation of the pre-trained text model while focusing optimization on the geometric modality, which is more diverse and less standardized. To enhance the generalization capability of our model, we apply lightweight augmentations to both modalities during training. For text, we randomly shuffle the order of GO terms before concatenating them to form captions, mitigating positional bias and encouraging semantic robustness. For point clouds, we apply geometric transformations including random jittering, rotation, and translation. These augmentations help the model become invariant to small perturbations in spatial configuration and text structure, improving cross-modal alignment under real-world variability.

\section{Experiments}

\subsection{Evaluation Tasks}

We design two evaluation tasks to assess the model's ability to generalize beyond the training set in a zero-shot setting. In both tasks, the model is not fine-tuned on evaluation data.

\minisection{Zero-shot retrieval on PDB.}  
Under this setting, we evaluate the model’s ability to retrieve functional annotations for unseen proteins. For each test point cloud sampled from a PDB entry, we construct a candidate pool of 100 GO-derived captions: one correct (positive) caption paired with the structure during pretraining, and 99 randomly selected negative captions. Cosine similarity scores are computed between the point cloud embedding and each caption embedding. We report Top-1 and Top-5 retrieval accuracy, where success is defined as the correct caption appearing among the top-$k$ ranked candidates.

\minisection{Cross-database zero-shot retrieval (EMDB $\rightarrow$ PDB).}  
To evaluate generalization across structural domains, we use EMDB-derived point clouds as query inputs. For each EMDB volume that has a corresponding PDB entry that is not exposed to the model during pretraining, we sample a point cloud and perform retrieval against the same candidate caption pool construction as above, using the GO annotation of the associated PDB entry as ground truth. Since EMDB structures are unseen during training, this setting tests the robustness of the learned embedding space under domain shift. Again Top-1 and Top-5 retrieval accuracy are recorded.
Following~\cite{nakamura2023daq,terashi2022residue}, we exclude poorly matched model-map pairs resulting from data collection inconsistencies between the two datasets. For more details, see~\cref{sec:corr}.  
This criterion excludes those pairs that have a sum of cross-correlation coefficients (CCC) and experimental-simulated map overlap (OVR) below 1.5, ensuring that only well-fitted structural models were retained for analysis. The calculation is done between the experimental map density from EMDB and the simulated map density generated from PDB structures using EMAN2's \texttt{e2pdb2mrc} function. The cross-correlation coefficients (CCC) are calculated as:
\begin{equation}
\text{CCC} = \frac{\sum_n p_{\text{exp}}(n) p_{\text{simu}}(n)}{\sqrt{\sum_n p_{\text{exp}}(n)^2 \sum_n p_{\text{simu}}(n)^2}}.
\end{equation}
Here, $p_{\text{exp}}(n)$ is the density of the experimental map at all points in the grid $n \in N$, where $N$ is the number of all grid point locations that have non-zero density in both the experimental and simulated map and $p_{\text{simu}}(n)$ is the density of the atomic model at the same point. We also calculate the overlap ratio (OVR) as:
\begin{equation}
\text{OVR} = \frac{\sum_n [(p_{\text{exp}}(n) \cdot p_{\text{simu}}(n)) > 0]}{\min\left(\sum_n [p_{\text{exp}}(n) > 0], \sum_n [p_{\text{simu}}(n) > 0]\right)}.
\end{equation}
Here, $p_{\text{exp}}(n)$ is the density of the experimental map at all points in the grid $n \in N$, where $N$ is the collection of all grid point locations, and $p_{\text{simu}}(n)$ is the density of the simulated map at the same point.

\subsection{Results}

\begin{table}
  \centering
  \begin{adjustbox}{width=0.5\linewidth}
  \begin{tabular}{@{}cc@{}}
    \toprule
    Top-1 acc. (\%)  & Top-5 acc. (\%)\\
    \midrule
    35.18 & 59.70 \\
    \bottomrule
  \end{tabular}
  \end{adjustbox}
  \vspace{-0.4em}
  \caption{Zero-shot Retrieval Results on PDB.}
  \vspace{-0.3em}
  \label{tab:pdb}
\end{table}

\begin{table}
  \centering
  \begin{adjustbox}{width=0.7\linewidth}
  \begin{tabular}{@{}ccc@{}}
    \toprule
    $\text{CCC}+\text{OVR}$ & Top-1 acc. (\%) & Top-5 acc. (\%) \\
    \midrule
    1.5 & 16.54 & 35.76 \\
    1.6 & 17.07 & 36.31 \\
    1.7 & 17.97 & 36.55 \\
    1.8 & 17.63 & 34.10 \\
    \bottomrule
  \end{tabular}
  \end{adjustbox}
  \vspace{-0.4em}
  \caption{Cross-database zero-shot retrieval results on EMDB.}
  \vspace{-1.0em}
  \label{tab:emdb}
\end{table}


Our findings show that aligning 3D point cloud representations of proteins with functional annotations through contrastive learning is promising for protein function retrieval. The model demonstrates promising zero-shot generalization, both within the PDB domain (see~\cref{tab:pdb}) and across domains using EMDB-derived structures (see~\cref{tab:emdb}). This indicates the potential of vision-language pretraining for biomolecular data, even when working with coarse, surface-level geometric input.
The use of point clouds—rather than full atomic coordinates—offers a practical abstraction akin to data obtainable from cryo-EM or predicted models, making it suitable for downstream function prediction where high-resolution detail is unavailable. Moreover, the ability to align these with symbolic GO terms suggests that foundation model concepts can be extended to structure-function modeling in molecular biology.



\section{Conclusion}

We present a CLIP-style contrastive learning framework that aligns 3D protein structures with GO-derived captions. Leveraging a large-scale dataset of protein–caption pairs, we show that point cloud–based representation can support zero-shot and cross-database protein function retrieval. This work demonstrates the potential of multimodal foundation models for protein-level understanding and lays the groundwork for future structure-to-function modeling, particularly for data from techniques like cryo-EM.

\minisection{Limitation discussion.} Despite the promising results, several limitations remain. Retrieval accuracy can be further improved, particularly for proteins with closely related functions. Additionally, assigning a single caption per protein restricts the model's ability to reflect functional diversity. In future work, we plan to enhance the model and training objective to boost retrieval precision. The framework is also extendable to other Gene Ontology categories, such as biological processes and cellular components, enabling more comprehensive structure–function modeling.

{
    \small
    \bibliographystyle{ieeenat_fullname}
    \bibliography{main}
}

\maketitlesupplementary

\section{EMDB-PDB Entry Correspondence}
\label{sec:corr}

To evaluate the generalization ability of the model across modalities, we perform zero-shot retrieval using EMDB entries, which are not seen by the model during training. Specifically, PDB entries corresponding to these EMDB entries were excluded from the training set, ensuring a true zero-shot setting even under the cross-database condition. 

However, database-linked EMDB–PDB pairs can vary in quality, with some exhibiting poor structural alignment due to resolution mismatch, fitting errors, or biological heterogeneity. To ensure reliable evaluation, we filter out ill-posed pairs using the sum of two alignment metrics: the cross-correlation coefficient (CCC) and the model–map overlap ratio (OVR)~\cite{nakamura2023daq, terashi2022residue}. Below, we present two representative examples—one excluded and one retained—to qualitatively support the effectiveness of this criterion.

\begin{figure}[t]
    \centering
    \includegraphics[width=\linewidth]{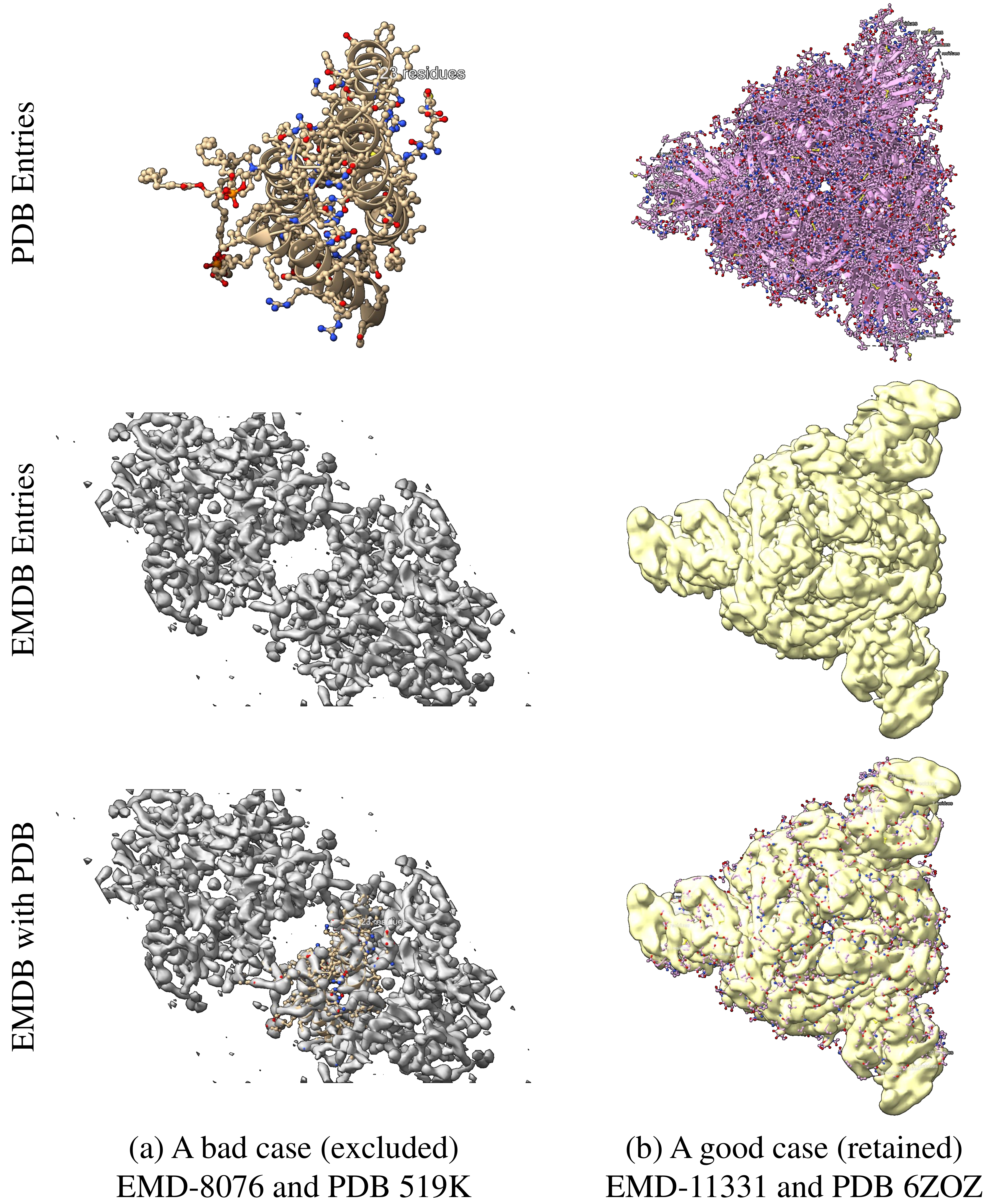}
    \caption{Bad case and good case examples. (a) A bad case example. $\text{CCC} = 0.13$, $\text{OVR} = 0.37$, $\text{CCC} + \text{OVR} = 0.50$. (b) A good case example. $\text{CCC} = 0.88$, $\text{OVR} = 0.88$, $\text{CCC} + \text{OVR} = 1.76$.}
    \vspace{-0.25em}
    \label{fig:case}
\end{figure}

\minisection{Bad case (excluded).} \cref{fig:case} (a) shows EMDB entry \texttt{EMD-8076} and its associated PDB model \texttt{5I9K}, which has a CCC of 0.13 and an OVR of 0.37, summing to 0.50. Although these entries are linked, the atomic model does not fit the density map well. As seen in the third row, the atomic model overlaps with only a small portion of the density map, indicating poor structural correspondence. The pairs similar to these examples are excluded from the cross-database evaluation set.

\minisection{Good case (retained).} \cref{fig:case} (b) presents \texttt{EMD-11331} and its corresponding PDB structure \texttt{6ZOZ}, with CCC and OVR values of 0.88 each (sum: 1.76). The atomic model and density map demonstrate strong geometric agreement, with the atomic structure fitting tightly within the experimental density map. The pairs with good correspondence similar to this, reflecting high-quality structure–map alignment are retained for cross-database zero-shot evaluation.

\section{Qualitative Retrieval Results}
\label{sec:qual}

\begin{figure}[H]
  \centering
  \includegraphics[width=0.65\linewidth]{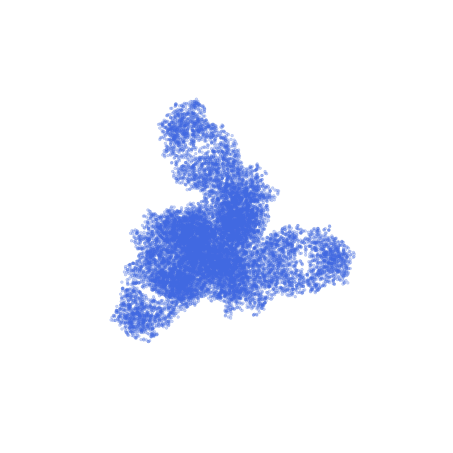}
  \vspace{-20pt}

  \small
  \begin{tabular}{p{0.95\linewidth}}
  \toprule
  \textbf{Ground-truth Caption:} identical protein binding, structural molecule activity, metal ion binding \\
  \textbf{Ground-truth Ranking in Retrieved Result:} 1/100 \\
  \midrule
  \textbf{Top-5 Retrieved Captions:} \\
  \textbf{\textcolor{RoyalBlue}{1. identical protein binding, structural molecule activity, metal ion binding}} \\
  2. structural molecule activity, identical protein binding \\
  3. identical protein binding, receptor ligand activity, structural constituent of virion, host cell surface receptor binding \\
  4. identical protein binding \\
  5. structural molecule activity \\
  \bottomrule
  \end{tabular}

  \caption{Retrieval result for query protein 5VJ6}
  \label{fig:5VJ6}
\end{figure}

\begin{figure}[H]
  \centering
  \includegraphics[width=0.65\linewidth]{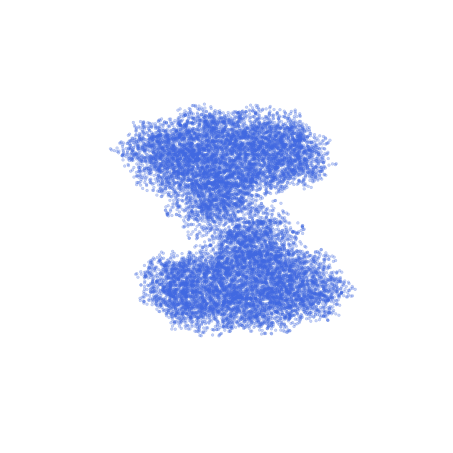}
  \vspace{-20pt}

  \small
  \begin{tabular}{p{0.95\linewidth}}
  \toprule
  \textbf{Ground-truth Caption:} serine C-palmitoyltransferase activity, pyridoxal phosphate binding \\
  \textbf{Ground-truth Ranking in Retrieved Result:} 4/100 \\
  \midrule
  \textbf{Top-5 Retrieved Captions:} \\
  1. oxalyl-CoA decarboxylase activity, thiamine pyrophosphate binding, identical protein binding, magnesium ion binding, ADP binding \\
  2. methylaspartate mutase activity, cobalamin binding, metal ion binding \\
  3. catalase activity, heme binding, nucleotide binding, metal ion binding \\
  \textbf{\textcolor{RoyalBlue}{4. serine C-palmitoyltransferase activity, pyridoxal phosphate binding}} \\
  5. structural molecule activity, metal ion binding \\
  \bottomrule
  \end{tabular}

  \caption{Retrieval result for query protein 7K0M}
  \label{fig:7K0M}
\end{figure}

\begin{figure}[H]
  \centering
  \includegraphics[width=0.65\linewidth]{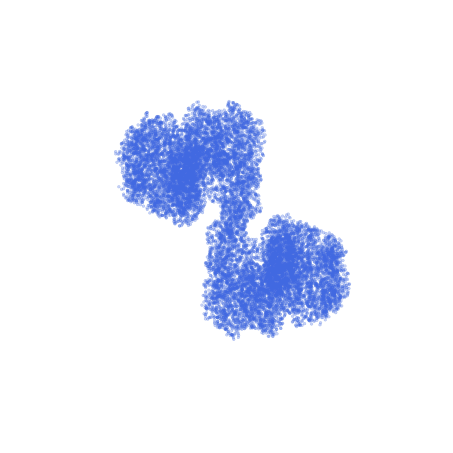}
  \vspace{-20pt}

  \small
  \begin{tabular}{p{0.95\linewidth}}
  \toprule
  \textbf{Ground-truth Caption:} cystathionine beta-synthase activity, nitric oxide binding, pyridoxal phosphate binding, nitrite reductase (NO-forming) activity, oxygen binding, ubiquitin protein ligase binding, carbon monoxide binding, modified amino acid binding, S-adenosyl-L-methionine binding, metal ion binding, heme binding, protein homodimerization activity \\
  \textbf{Ground-truth Ranking in Retrieved Result:} 60/100 \\
  \midrule
  \textbf{Top-5 Retrieved Captions:} \\
  1. carbohydrate binding, glycosyltransferase activity \\
  2. methionine-tRNA ligase activity, ATP binding \\
  3. K63-linked polyubiquitin modification-dependent protein binding, cysteine-type deubiquitinase activity, ubiquitin protein ligase binding, protein tag activity, metal ion binding, structural constituent of ribosome \\
  4. hydrolase activity, hydrolyzing O-glycosyl compounds \\
  5. NAD+-protein-arginine ADP-ribosyltransferase activity, nucleotide binding \\
  \bottomrule
  \end{tabular}

  \caption{Retrieval result for query protein 8S5K}
  \label{fig:8S5K}
\end{figure}

\begin{figure}[H]
  \centering
  \includegraphics[width=0.65\linewidth]{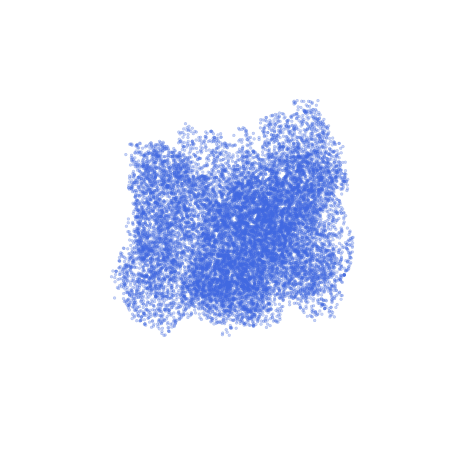}
  \vspace{-20pt}

  \small
  \begin{tabular}{p{0.95\linewidth}}
  \toprule
  \textbf{Ground-truth Caption:} symporter activity \\
  \textbf{Ground-truth Ranking in Retrieved Result:} 1/100 \\
  \midrule
  \textbf{Top-5 Retrieved Captions:} \\
  \textbf{\textcolor{RoyalBlue}{1. symporter activity}}\\
  2. DNA binding \\
  3. mitochondrial promoter sequence-specific DNA binding, mitochondrial transcription factor activity \\
  4. monoatomic cation transmembrane transporter activity, metal ion binding \\
  5. RNA polymerase II cis-regulatory region sequence-specific DNA binding, mediator complex binding \\
  \bottomrule
  \end{tabular}

  \caption{Retrieval result for query protein 6XWR}
  \label{fig:6XWR}
\end{figure}

\begin{figure}[H]
  \centering
  \includegraphics[width=0.42\linewidth]{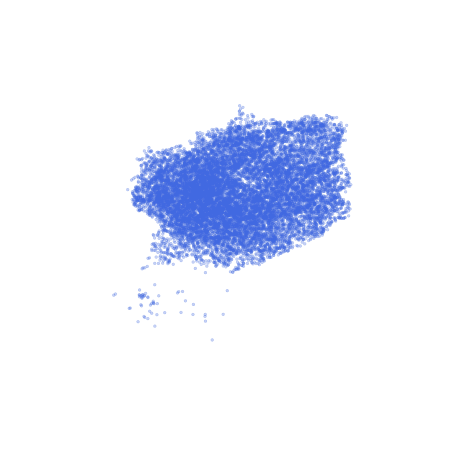}
  \vspace{-18pt}

  \small
  \begin{tabular}{p{0.95\linewidth}}
  \toprule
  \textbf{Ground-truth Caption:} endonuclease activity, RNA binding, RNA-dependent RNA polymerase activity, metal ion binding, nucleotide binding \\
  \textbf{Corresponding PDB Entry:} 7NI0 \\
  \textbf{Ground-truth Ranking in Retrieved Result:} 1/100 \\
  \midrule
  \textbf{Top-5 Retrieved Captions:} \\
  \textbf{\textcolor{RoyalBlue}{1. endonuclease activity, RNA binding, RNA-dependent RNA polymerase activity, metal ion binding, nucleotide binding}} \\
  2. peptidase activity, proteasome binding, protein tag activity, hydrolase activity, acting on carbon-nitrogen (but not peptide) bonds, in linear amides, metal ion binding, ATP binding \\
  3. zinc ion binding, ribonucleoside binding, magnesium ion binding, protein dimerization activity, DNA binding \\
  4. proton-transporting ATP synthase activity, rotational mechanism, proton-transporting ATPase activity, rotational mechanism, ATP hydrolysis activity, ATP binding, lipid binding, ADP binding \\
  5. aminopeptidase activity, serine-type peptidase activity, dipeptidyl-peptidase activity \\
  \bottomrule
  \end{tabular}
    
  \caption{Retrieval result for query protein EMD-12348}
  \label{fig:EMD-12348}
\end{figure}

\begin{figure}[H]
  \centering
  \includegraphics[width=0.42\linewidth]{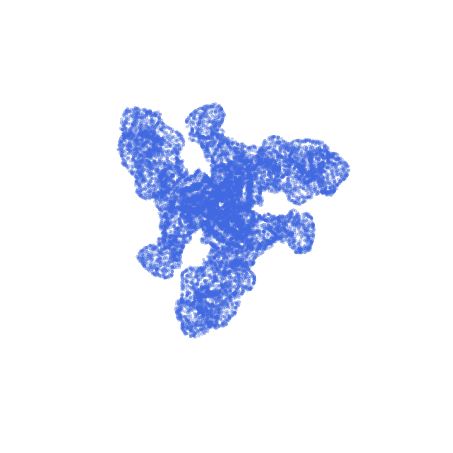}
  \vspace{-18pt}

  \small
  \begin{tabular}{p{0.95\linewidth}}
  \toprule
  \textbf{Ground-truth Caption:} structural molecule activity, identical protein binding \\
  \textbf{Corresponding PDB Entry:} 6E5P \\
  \textbf{Ground-truth Ranking in Retrieved Result:} 21/100 \\
  \midrule
  \textbf{Top-5 Retrieved Captions:} \\
  1. RNA endonuclease activity, DNA binding \\
  2. RNA endonuclease activity, RNA binding \\
  3. retinol binding, chemokine activity, complement binding, endopeptidase inhibitor activity, protein-containing complex binding \\
  4. oxalate decarboxylase activity, metal ion binding \\
  5. metal ion binding \\
  \bottomrule
  \end{tabular}
    
  \caption{Retrieval result for query protein EMD-8981}
  \label{fig:EMD-8981}
\end{figure}

\begin{figure}[H]
  \centering
  \includegraphics[width=0.42\linewidth]{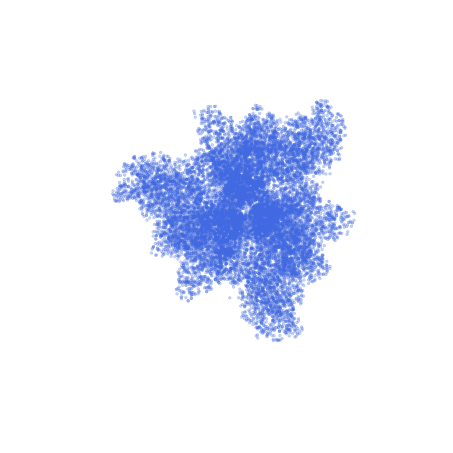}
  \vspace{-18pt}

  \small
  \begin{tabular}{p{0.95\linewidth}}
  \toprule
  \textbf{Ground-truth Caption:} identical protein binding, receptor ligand activity, structural constituent of virion, host cell surface receptor binding \\
  \textbf{Corresponding PDB Entry:} 7JVC \\
  \textbf{Ground-truth Ranking in Retrieved Result:} 1/100 \\
  \midrule
  \textbf{Top-5 Retrieved Captions:} \\
  \textbf{\textcolor{RoyalBlue}{1. identical protein binding, receptor ligand activity, structural constituent of virion, host cell surface receptor binding}} \\
  2. glycine-gated chloride ion channel activity, transmembrane signaling receptor activity, metal ion binding, extracellularly glycine-gated chloride channel activity, glycine binding \\
  3. maltose binding, BH3 domain binding, channel activity, protein heterodimerization activity, carbohydrate transmembrane transporter activity, protein transmembrane transporter activity \\
  4. hydrolase activity, four-way junction helicase activity, DNA binding, ATP binding \\
  5. toxin activity, calcium channel regulator activity \\
  \bottomrule
  \end{tabular}

  \caption{Retrieval result for query protein EMD-22506}
  \label{fig:EMD-22056}
\end{figure}

\begin{figure}[H]
  \centering
  \includegraphics[width=0.42\linewidth]{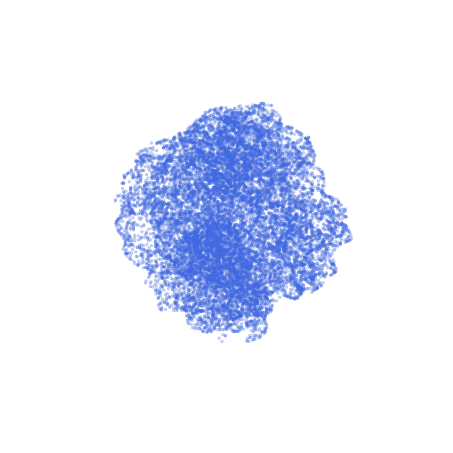}
  \vspace{-18pt}

  \small
  \begin{tabular}{p{0.95\linewidth}}
  \toprule
  \textbf{Ground-truth Caption:} molecular condensate scaffold activity, GTP binding, double-stranded DNA binding, phosphatidylinositol-4,5-bisphosphate binding, poly-ADP-D-ribose modification-dependent protein binding, protein heterodimerization activity, RNA binding, structural constituent of chromatin, metal ion binding, nucleosome binding, ATP binding, 2',3'-cyclic GMP-AMP synthase activity, enzyme binding, protein homodimerization activity \\
  \textbf{Corresponding PDB Entry:} 6XJD \\
  \textbf{Ground-truth Ranking in Retrieved Result:} 56/100 \\
  \midrule
  \textbf{Top-5 Retrieved Captions:} \\
  1. single-stranded DNA binding \\
  2. citrate (Si)-synthase activity \\
  3. phosphoenolpyruvate carboxylase activity \\
  4. transcription factor binding, DNA binding, metal ion binding \\
  5. serine-type endopeptidase inhibitor activity \\
  \bottomrule
  \end{tabular}

  \caption{Retrieval result for query protein EMD-22206}
  \label{fig:EMD-22206}
\end{figure}

To provide a deeper insight into model behavior, we present qualitative retrieval examples from both the PDB and EMDB test sets. Figs.~\ref{fig:5VJ6}–\ref{fig:6XWR} show zero-shot results on the PDB test set, while Figs.~\ref{fig:EMD-12348}–\ref{fig:EMD-22206} illustrate cross-database zero-shot retrieval on EMDB entries, where no corresponding PDB structure was seen during training. Each example includes a point cloud visualization of the query protein, its corresponding ground-truth caption, the ranking position of the ground truth among the 100 candidates, and the top-5 retrieved captions sorted by cosine similarity. For cross-database zero-shot retrieval results, the PDB entry ID corresponding the the query protein from EMDB is included.

These examples highlight a spectrum of retrieval outcomes. In high-performing cases (e.g., Figs.~\ref{fig:5VJ6},~\ref{fig:6XWR},~\ref{fig:EMD-12348},~\ref{fig:EMD-22056}), the ground-truth caption appears at the top of the ranked list or exhibits strong lexical and semantic overlap with retrieved candidates, indicating that the model successfully captures relevant structural features. In contrast, other cases (e.g., Figs.~\ref{fig:8S5K},~\ref{fig:EMD-8981},~\ref{fig:EMD-22206}) reveal current limitations, where retrieved captions deviate meaningfully from the ground truth or rank it substantially lower.These unsatisfactory cases often involve compositionally complex or lengthy functional descriptions that may surpass the text encoder's ability to effectively represent all terms, particularly when numerous specific GO terms are concatenated.

Together, these qualitative examples complement the quantitative results by offering a more nuanced view of the generalization capacity of the model and zero-shot retrieval behavior in both within-domain (PDB entries) and cross-database (EMDB entries) settings. Ground-truth captions that appear in the top-5 predictions are highlighted in \textcolor{RoyalBlue}{blue} for visual reference.

\clearpage

\vspace{-50pt}


\vspace{-30pt}

\clearpage

\end{document}